# A PLANAR VACUUM DIODE WITH ELECTRONS POSSESSING VELOCITY DISTRIBUTION FUNCTION

**Dimitar G. Stoyanov**

Department of Physics, Sofia Technical University, Sliven Branch

59, Bourgasko Shaussee Blvd, 8800 Sliven, Republic of Bulgaria

E-mail: dgstoyanov@abv.bg

**Abstract**

*The current-voltage characteristic of a planar vacuum diode with electrons possessing velocity distribution function during an emission from the cathode (electron beam) is theoretically considered in this article. The electron beam movement in the volume is examined. An integral-differential equation for the electric field potential is got in a correct form. A current-voltage characteristic of the diode for a model function of the velocity distribution of the emitted electrons is obtained.*

## 1. INTRODUCTION

The theory of vacuum diode with thermionic emission from the cathode has been developed since the beginning of the XX century by Child, Langmuire et. all [1-4].

Later some attempts for set of modifications have been done aiming the setting of the theory in accordance with the certain problem (e.g. see [5-7]). During the theory construction some fundamental suggestions and approximations had been done [1-3], which were keeping in the following developments, but undergoing serious criticism [8]. This is the reason all next modifications of the theory to be partial and to have no common theoretical significance.

The objective of the present article is the giving of a correct formulation of the equation describing the electric field in a planar vacuum diode in a stationary case, when from the cathode are emitting electrons possessing a velocity distribution.

## 2. A PLANAR DIODE WITH ELECTRONS POSSESSING A VELOCITY DISTRIBUTION FUNCTION

### 2.1 Geometry and basic equations

A diode is a system of two parallel endless metallic planes situated apart in a distance **d**, and put in high vacuum. The spatial symmetry is such that we can choose a Cartesian coordinate system whose axes OX and OY lay in the metallic plane performing as a cathode, and an axis OZ perpendicular to the cathode. Thus, at such chosen coordinate system the cathode has a coordinate $z = 0$, while the anode has a coordinate $z = d$.

The geometry of the problem is such that all parameters of the electric field and the velocities of the electrons depend merely on the distance to the cathode **z**.

The cathode ($z = 0$) emits electrons (with electrical charge **−q**) possessing a velocity distribution function. In this case the velocity vectors are directed towards the positive direction of the axis OZ, and the vector of current density is pointed from the anode to the cathode.

For the function of velocity distribution function of the current density of the electrons emitted from the cathode we will put

$$dj = j_0 . f(v) . dv, \qquad (1)$$

where  - **dj** is the current density of the emitted electrons with velocity in the
interval $(v, v + dv)$;
- $j_0$ is the current density of all emitted electrons;
- **v** is the velocity of the electrons;
- $f(v)$ is a function of velocity distribution of the current density.

In common case we suppose that the electrons emitted from the cathode have velocities $v \in [0, \infty)$. Therefore the condition upon the cathode for the normalization of the function of distribution of the emitted electrons is of the following form

$$\int_0^\infty f(v) . dv = 1. \qquad (2)$$

Whence the current density of all emitted electrons $j_0$ coincides with density of the current of saturation of the vacuum diode [8].

The form of the function of distribution upon the cathode depends on the nature and the emission mechanism of the electrons from the cathode, and it is external for the considered problem. We accept that this distribution is assigned.

We choose to represent the emitted electrons by (1) because in the stationary case in the diode volume is executed the following condition

$$\nabla \cdot \vec{j} = 0. \tag{3}$$

The electrons emitted by the cathode get onto the volume between both electrodes, thus forming the potential of the electric field $\varphi(z)$. The equipotential surfaces of the electric field are planes parallel to the plane XOY. We put the potential of the electric field upon the cathode is $\varphi(0) = 0$. The anode ($z = d$) is found at potential $\varphi(d) = U$.

According to [1-3, 9] the potential of the electric field in the volume between the anode and the cathode is defined from the equation of Poisson

$$\nabla^2 \varphi(z) = -\frac{q \cdot n_e(z)}{\varepsilon_0}, \tag{4}$$

where - $n_e(z)$ is the density of the electrons;

- $\varepsilon_0$ is the permittivity of free space.

The function of the potential of the electric field $\varphi(z)$ possesses a minimum [4, 8] in the point with coordinate $z_m$ ($0 \leq z_m \leq d$) and magnitude $\varphi_m = \varphi(z = z_m) < 0$.

For the determination of the current density in the volume between the electrodes the movement of the electrons in the volume is necessary to be considered.

### 2.2. A trajectory of the electrons in the phase space

The model running of the change of the electrons velocity $v(z)$ in the volume of the diode is represented in Figure 1. The figure illustrates the fact that the electrons emitted by the cathode penetrate in the diode volume and their velocity at first (close to the cathode) begins to decrease. The reason for this is the negative potential of the electric field near to the cathode [4, 8].

In a point with coordinate $z$ each electron emitted by the cathode with velocity $v(0)$ has a moment velocity $v(z)$ that is determined by the formula [1-3, 8, 9]

$$v(z) = \pm\sqrt{v(0)^2 - \frac{2.q.\varphi(z)}{m}}, \qquad (5)$$

where **m** is the electron mass.

Let at first examine the region round the cathode, i.e. $z \in [0, z_m]$.

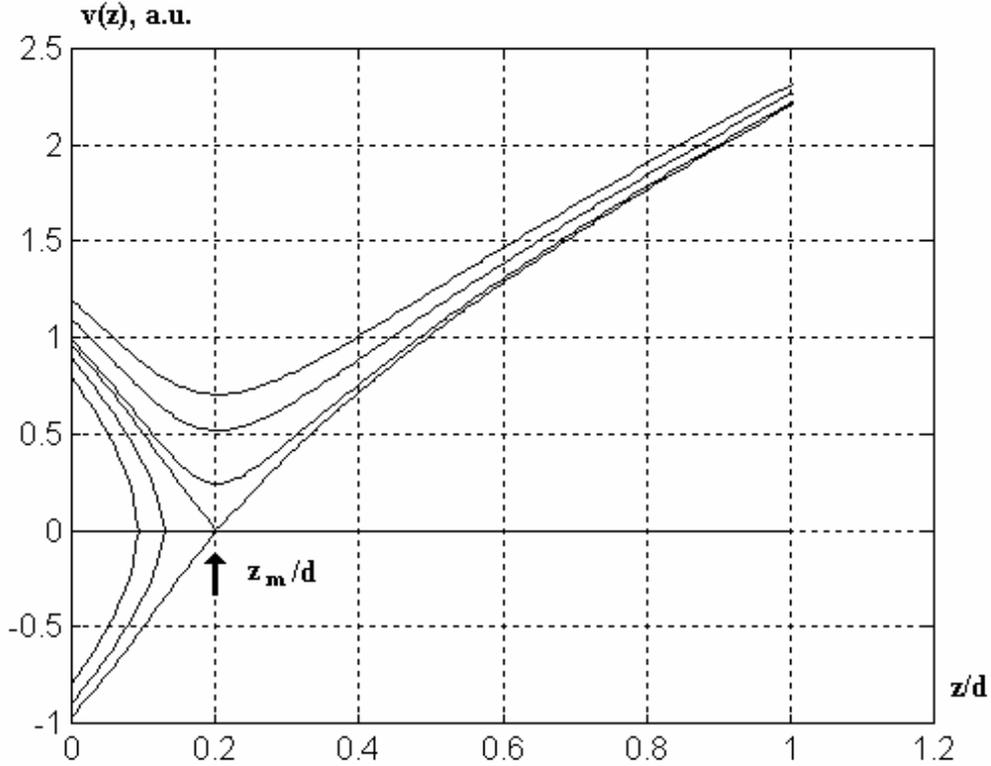

Figure 1. A model changing of the velocity of the electrons in the diode volume.

1) In this point electrons with a moment velocity of no value will be available. These are the electrons emitted from the cathode with velocity $v_\Phi$.

$$v_\Phi = \sqrt{\frac{2.q.\varphi(z)}{m}}. \qquad (6)$$

At the next moments these electrons will begin to move towards the cathode.

2) The electrons emitted from the cathode with velocity lower than $v_\Phi$ cannot reach at all this point. Somewhere between the cathode and this point their velocity is nullified itself, and they have begun to move in direction to the cathode.

3) The electrons emitted from the cathode with velocity higher than $v_\Phi$ will continue their movement in direction to the anode.

But the anode will be reached by only these electrons which possess during the emission a velocity higher than $v_m$

$$v_m = \sqrt{\frac{2.q.\varphi_m}{m}}. \tag{7}$$

The electrons possessing during the emission a velocity $v(0) \in (v_\Phi, v_m)$ will pass through the point with coordinate $z$. But because of that their velocity is not sufficient for the overcoming of the potential barrier at $z_m$, somewhere in the interval $(z, z_m)$ their velocity will be nullified, and they will begin to move to the cathode.

That is why we have to expect the presence of two groups of electrons in each point with coordinate $z < z_m$:

- Electrons moving to the anode. These are electrons emitted from the cathode with velocity higher than $v_\Phi$.
- Electrons coming back to the cathode. These are electrons emitted from the cathode with velocity $(v_\Phi, v_m)$ passing across the point with coordinate $z$ during their movement to the anode, but which have returned because of a deficiency of velocity.

In the region where $z \in [z_m, d]$ the potential of the electric field is an increasing function and nothing can stop the electrons. But here we can find only electrons emitted by the cathode with velocity higher than $v_m$ (these are the electrons which have overcome the potential barrier at $z_m$). These electrons reach the anode and form a current density through the diode

$$j = j_0 . \int_{v_m}^{\infty} f(v(0)).dv(0). \tag{8}$$

**2.3. Equation of the electric field**

For the obtaining of the density of the electrons lying in point $z$ we use the continuity of the current density (3) and integrating by the velocity with which the electrons are emitted we get

$$n_e(z) = \frac{j_0}{q} \cdot \int_{v_m}^{\infty} \frac{f(v(0)).dv(0)}{v(z)} + 2 \cdot \frac{j_0}{q} \cdot \int_{v_\varphi}^{v_m} \frac{f(v(0)).dv(0)}{v(z)}, \quad z \leq z_m, \tag{9a}$$

$$n_e(z) = \frac{j_0}{q} \cdot \int_{v_m}^{\infty} \frac{f(v(0)).dv(0)}{v(z)}, \quad z \geq z_m. \tag{9b}$$

The integration in (9a) and (9b) is done by the velocity with which the electrons are emitted from the cathode. This allows the obtaining of results without getting of the function of the electrons distribution in each spatial point.

The first term in the right side of (9a) gives the density of those electrons which will overcome the potential barrier at $z_m$.

The second term gives the density of those electrons which will reach points from the interval $(z, z_m)$ and will return again. Therefore they will come twice across the point with coordinate $z$. Whence is the coefficient 2. The lower integral limit of the second term of (9a) has the sense of an initial velocity of these electrons (6), which getting in point $z$ will have a moment velocity of no value.

If in (9a) and (9b) replace (5), and the so got phrases replace in (4) we will obtain an integral-differential equation for the potential of the electric field in the vacuum diode.

After scaling of the energy (we introduce a scale of the energy $E_1$) and changing the variables the equation is produced in dimensionless units in the following form

$$\frac{d^2\Phi}{dx^2} = J_0 \cdot \int_{\eta_m}^{\infty} \frac{f(\eta).d\eta}{\sqrt{\eta^2 + \Phi}} + 2.J_0 \cdot \int_{\eta_\Phi}^{\eta_m} \frac{f(\eta).d\eta}{\sqrt{\eta^2 + \Phi}}, \quad x \leq x_m, \tag{10a}$$

$$\frac{d^2\Phi}{dx^2} = J_0 \cdot \int_{\eta_m}^{\infty} \frac{f(\eta).d\eta}{\sqrt{\eta^2 + \Phi}}, \quad x \geq x_m. \tag{10b}$$

where  - $x = \frac{z}{d}, \quad x \in [0, 1];$

- $\Phi(x) = -\frac{q.\varphi(z)}{E_1};$

- $\Phi d = -\dfrac{q.U}{E_1}$;

- $\Phi m = -\dfrac{q.\varphi_m}{E_1}$;

- $J_0 = \dfrac{q.j_0.d^2}{E_1.\varepsilon_0} \cdot \sqrt{\dfrac{m}{2.E_1}}$;

- $\eta = \sqrt{\dfrac{m.v(0)^2}{2.E_1}}$;

- $\eta_m = \sqrt{-\Phi_m}$;

- $\eta_\Phi = \sqrt{-\Phi(z)}$.

On the style of the new variables (8) gets the form

$$J = J_0 \cdot \int_{\eta_m}^{\infty} f(\eta).d\eta. \tag{11}$$

where $\quad J = \dfrac{q.j.d^2}{E_1.\varepsilon_0} \cdot \sqrt{\dfrac{m}{2.E_1}}.$

Thus we get the relation between the density of the current flowing through the diode $J$, the density of the current of saturation of the diode $J_0$, and the minimum of the electric field potential $\Phi_m$ in dimensionless units.

The first integration of (10a) and (10b) in common case is easily done, and as a result is got the following

$$\left(\dfrac{d\Phi}{dx}\right)^2 = 4.J_0 \cdot \int_{\eta_m}^{\infty} d\eta.f(\eta).\left(\sqrt{\eta^2 + \Phi} - \sqrt{\eta^2 + \Phi_m}\right) +$$

$$+ 8.J_0 \cdot \int_{\eta_\Phi}^{\eta_m} d\eta.f(\eta).\sqrt{\eta^2 + \Phi}, \qquad x \leq x_m. \tag{12a}$$

$$\left(\frac{d\Phi}{dx}\right)^2 = 4.J_0 \cdot \int_{\eta_m}^{\infty} d\eta.f(\eta).\left(\sqrt{\eta^2 + \Phi} - \sqrt{\eta^2 + \Phi_m}\right), \quad x \geq x_m. \tag{12b}$$

During the obtaining of (12a) and (12b) is taken into account that at $x = x_m$ the function of the potential $\Phi(x)$ has a minimum, in which

$$\Phi(x_m) = \Phi_m, \tag{13a}$$

$$\left.\frac{d\Phi}{dx}\right|_{x=x_m} = 0. \tag{13b}$$

The solving of (12a) and (12b) farther on requires a concretization of the form of the current density distribution and the corresponding integration. During this new integration is necessary to be read the fact that $\Phi(x = 0) = 0$ and $\Phi(x = 1) = \Phi d$.

Therefore the problem for the determination of the current-voltage characteristic of the vacuum diode with electrons velocities distribution is theoretically solved.

## 2.4. A model function of distribution

In the present article as a model is accepted the following distribution [9]

$$f(\eta) = A.\eta.\exp\left[-a.(\eta - 1)^2\right], \tag{14}$$

where - **A** is a constant;
- **a** is a parameter of distribution.

The constant **A** presented in (14) is determined by the condition for the normalization of the function of the distribution of the emitted electrons (2).

This distribution (14) has the meaning of a beam of electrons with dimensionless average velocity $\eta = 1$ and Maxwell's distribution of electrons round this average value with parameter **a** (a dimensionless temperature of the electrons $\theta = 1/a$). In this case we suppose that the parameter **a** has a value bigger than 50. Using such distribution we may describe a beam of electrons with "low" electron temperature.

The determination of the current-voltage characteristic of the diode with distribution (14) is done numerically following the next procedure:

- at assigned $J$ and $J_0$ using (14) from (11) $\eta_m$ and $\Phi_m$ are determined;
- using $J_0$ and $\Phi_m$ yet obtained we calculate (12a) and (12b);
- by numerical integration the potential $\Phi d = \Phi(x = 1)$ is got.

The calculations are done using the program product MATLAB for two values of the parameter $a$, $a = 100$ and $a = 900$. The results from the calculations of the current-voltage characteristic of the diode for $J_0 = 16$ with distribution (14) and for both value of parameter $a$ are represented in Figure 2.

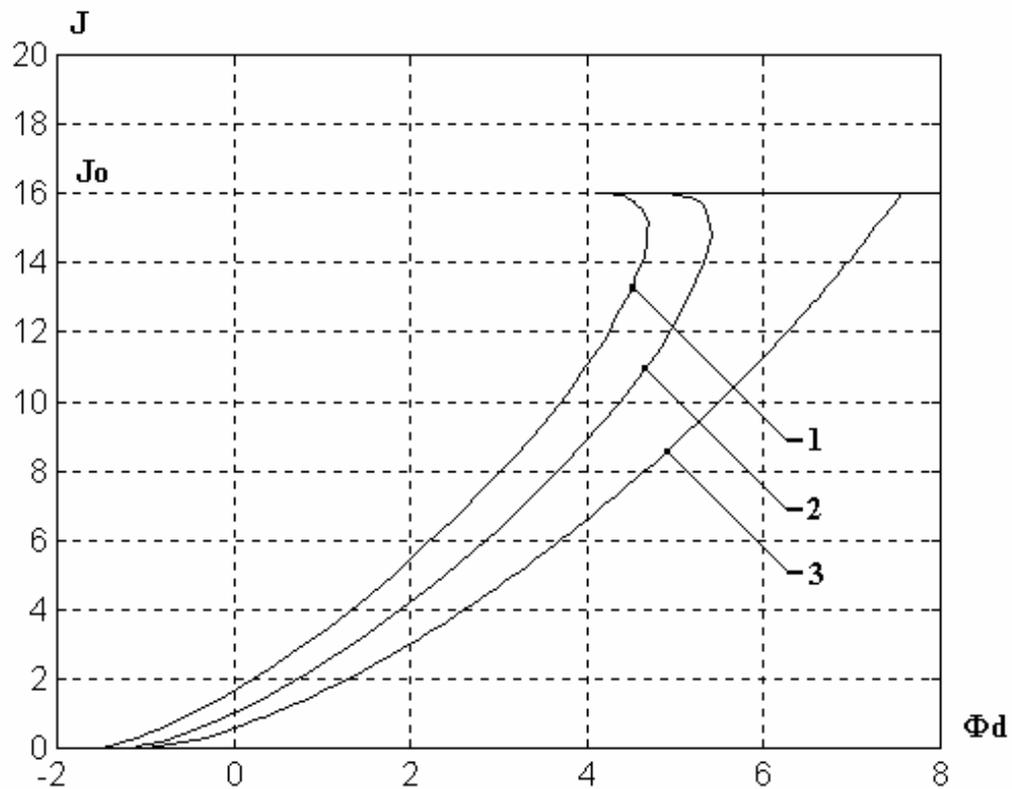

Figure 2. A current-voltage characteristic of a planar vacuum diode for $J_0 = 16$ with distribution (14) at the following values of the parameter a: curve 1, $a = 100$; curve 2, $a = 900$; curve 3, $a = \infty$ [8].

In the same figure for a comparison is shown a solution obtained in [8], which on the meaning of (14) is corresponding to $a = \infty$ (the monoenergetic beam of electrons is a beam of electrons with electron temperature with no value). From the figure is evident that there is no contradiction between both results. An expected result that the current-voltage characteristic of the vacuum diode with beam electrons possesses a section with **S**-instability at low electron temperature is seen in this figure.

## 3. CONCLUSION

Finally it can be summarized that integral-differential equations are obtained in the correct form, which are describing the electric field potential in the volume of vacuum diode with electrons possessing a velocity distribution. The taking into account of the returned to the cathode electrons and the application of a function of distribution in common form lead to results, which could be used for a planar diode with cathode thermionic emission and a planar vacuum photocell.

For a model distribution a current-voltage characteristic is obtained, and the presence of a section with S-instability for a beam of electrons with low electron temperature is ascertained.

**REFERENCES**

[1] Child C.D., Discharge from Hot CaO, *Phys. Rev.,* **32**, 492 (1911).

[2] Langmuire I., The Effect of space Charge and Residual Gases on Thermionic Currents in High Vacuum, *Phys. Rev.*, **2**, 450 (1913).

[3] Langmuire I., The Effect of space Charge and Initial velocities on the Potential Distribution and Thermionic Current between Parallel Plane Electrodes, *Phys. Rev.,* **21**, 419 (1923).

[4] Schottky W Cold and Hot Electron Discharges (in German), Z. Physik., **14**, 63 (1923).

[5] Luginsland J W, Y Y Lau and R M Gilgenbach *Phys. Rev. Let*. **77**, 4668 (1996);

Luginsland J W, YY Lau, R Umstattd and J J Watrous *Phys. Plasmas* **9,** 2371 (2002)

[6] Lau Y Y , *Phys. Rev. Lett.* **87**, 278301 (2001)

[7] Kostov K G, J J Barroso, *Phys. Plasmas* **9**, 1039 (2002)

[8] Stoyanov D G  Planar Vacuum Diode With Monoenergetic Electrons, *J. Appl. Electromagnetism*, V 8, No1,2 , June 2006, Dec. 2006, pp.35-48.

[9] Y. P. Raizer, *Gas Discharge Physics*, Berlin: Springer-Verlag, 1991.